\documentclass[12pt]{iopart}

\usepackage{graphicx}
\usepackage{dcolumn}
\usepackage{caption}

\begin{document}

\title[Energy and centrality dependence of particle production at very low $p_{T}$ ]{Energy and centrality dependence of particle production at very low transverse momenta in Au+Au collisions}

\author{ 
B.Alver$^4$,
B.B.Back$^1$,
M.D.Baker$^2$,
M.Ballintijn$^4$,
D.S.Barton$^2$,
R.R.Betts$^6$,
A.A.Bickley$^7$,
R.Bindel$^7$,
W.Busza$^4$,
A.Carroll$^2$,
Z.Chai$^2$,
V.Chetluru$^6$,
M.P.Decowski$^4$,
E.Garc\'{\i}a$^6$,
T.Gburek$^3$,
N.George$^2$,
K.Gulbrandsen$^4$,
C.Halliwell$^6$,
J.Hamblen$^8$,
I.Harnarine$^6$,
M.Hauer$^2$,
C.Henderson$^4$,
D.J.Hofman$^6$,
R.S.Hollis$^6$,
R.Ho\l y\'{n}ski$^3$,
B.Holzman$^2$,
A.Iordanova$^6$,
E.Johnson$^8$,
J.L.Kane$^4$,
N.Khan$^8$,
P.Kulinich$^4$,
C.M.Kuo$^5$,
W.Li$^4$,
W.T.Lin$^5$,
C.Loizides$^4$,
S.Manly$^8$,
A.C.Mignerey$^7$,
R.Nouicer$^2$,
A.Olszewski$^3$,
R.Pak$^2$,
C.Reed$^4$,
E.Richardson$^7$,
C.Roland$^4$,
G.Roland$^4$,
J.Sagerer$^6$,
H.Seals$^2$,
I.Sedykh$^2$,
C.E.Smith$^6$,
M.A.Stankiewicz$^2$,
P.Steinberg$^2$,
G.S.F.Stephans$^4$,
A.Sukhanov$^2$,
A.Szostak$^2$,
M.B.Tonjes$^7$,
A.Trzupek$^3$,
C.Vale$^4$,
G.J.van~Nieuwenhuizen$^4$,
S.S.Vaurynovich$^4$,
R.Verdier$^4$,
G.I.Veres$^4$,
P.Walters$^8$,
E.Wenger$^4$,
D.Willhelm$^7$,
F.L.H.Wolfs$^8$,
B.Wosiek$^3$,
K.Wo\'{z}niak$^3$,
S.Wyngaardt$^2$,
B.Wys\l ouch$^4$\\
\vspace{3mm}
\small
%
%
%
%
$^1$~Argonne National Laboratory, Argonne, IL 60439-4843, USA\\
$^2$~Brookhaven National Laboratory, Upton, NY 11973-5000, USA\\
$^3$~Institute of Nuclear Physics PAN, Krak\'{o}w, Poland\\
$^4$~Massachusetts Institute of Technology, Cambridge, MA 02139-4307, USA\\
$^5$~National Central University, Chung-Li, Taiwan\\
$^6$~University of Illinois at Chicago, Chicago, IL 60607-7059, USA\\
$^7$~University of Maryland, College Park, MD 20742, USA\\
$^8$~University of Rochester, Rochester, NY 14627, USA\\
}

\begin{abstract}
The PHOBOS experiment at RHIC has the unique capability of measuring particle production
at very low transverse momenta. New results on low-transverse momentum invariant yields 
of $\pi^{\pm}$ , $K^{\pm}$ and ($p+\bar{p}$) in 200 GeV Au+Au collisions are 
presented as a function of the collision centrality for the 50\% most central events.
In contrast to the results from d+Au collisions, no $m_{T}$  
scaling is observed in the very low $p_{T}$ region.
The low transverse momentum yields agree with extrapolations from intermediate transverse momentum measurements. 
For all collision centralities a flattening of the transverse momentum spectra is observed, consistent with 
a rapid transverse expansion of the system. 
\end{abstract}

\begin{figure}[h!]

\begin{minipage}[t]{7.7cm}
\includegraphics[width=7.7cm]{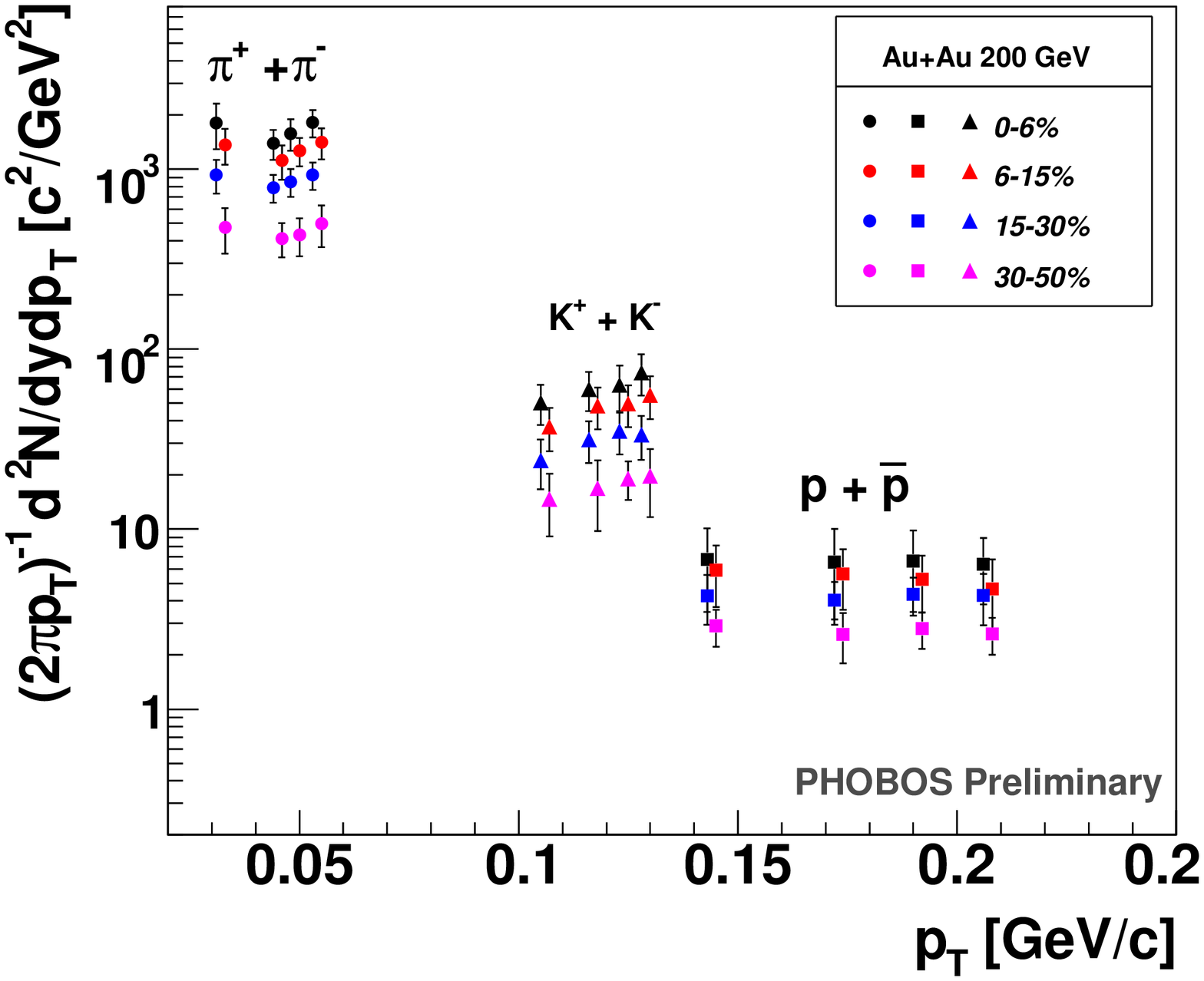}
\vspace{-1.0cm} 
\caption{ Identified particle spectra at very low $p_{T}$ near mid-rapidity in Au+Au collisions at 
          $\sqrt{s_{NN}}$~=~200~GeV. Combined systematic and statistical errors are shown. }
\label{spect200}
\end{minipage}
\hfill
\begin{minipage}[t]{7.7cm}
\includegraphics[width=7.7cm]{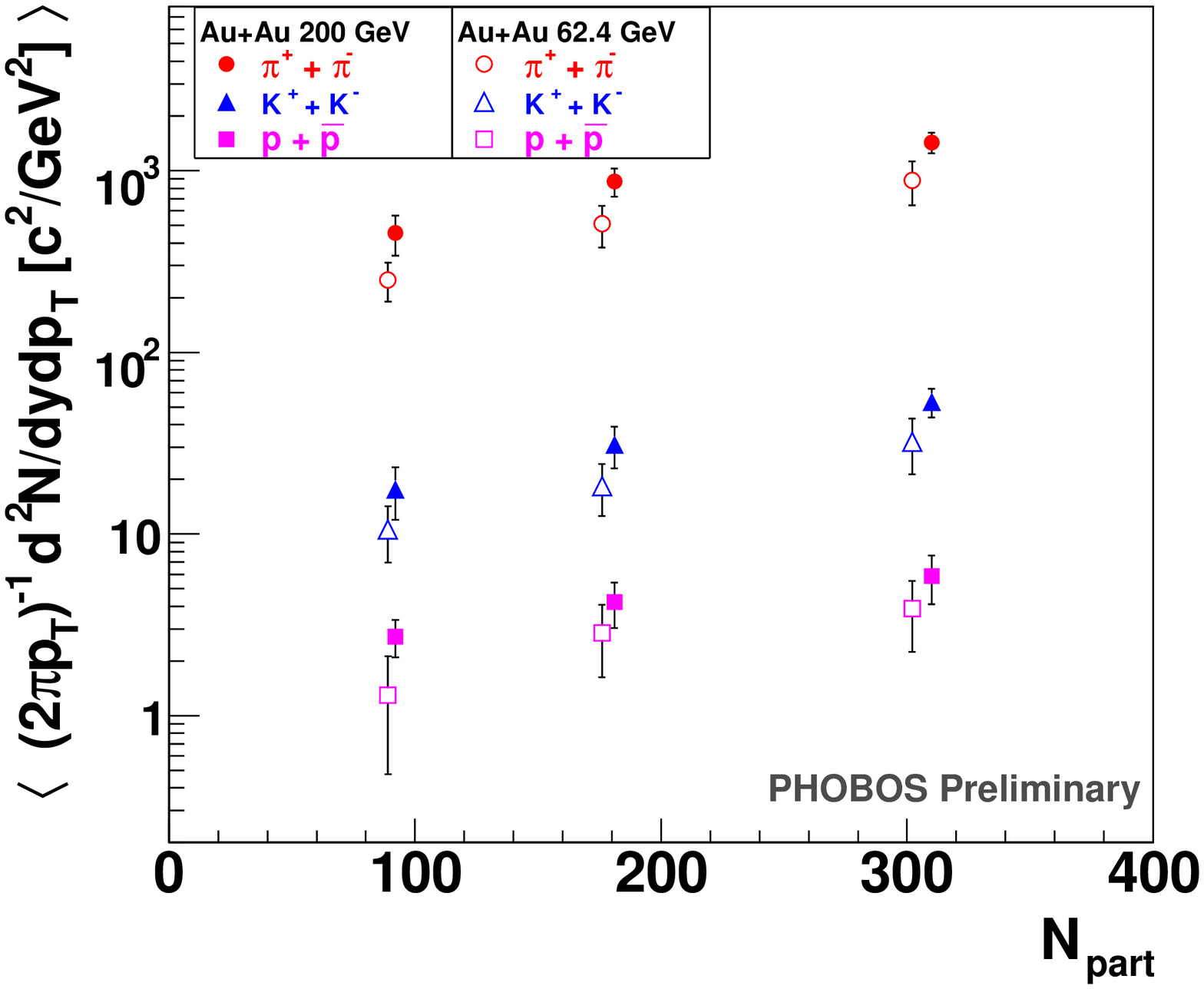} 
\vspace{-1.0cm}
\caption{ Invariant yields for Au+Au collisions at 200 GeV (solid~symbols) and 62.4 GeV (open~symbols) 
          as a function of centrality. The yields are averaged over $p_{T}$.}
\label{comp200_624}
\end{minipage}

\end{figure}

The PHOBOS detector at the Relativistic Heavy Ion  
Collider (RHIC) at Brookhaven National Laboratory has been used to  
measure low-transverse momentum invariant yields for $\pi^{\pm}$~,~$K^{\pm}$ and ($p+\bar{p}$) 
produced in collisions of gold nuclei at a center-of-mass 
energy per nucleon pair of 200 GeV (Fig.~\ref{spect200}).
The analysis is based on a large statistics dataset that allows the data to be split in four different centrality classes 0-6\%, 6-15\%, 15-30\% and 30-50\% most central events instead of the one centrality class that was available for the first 200 GeV analysis \cite{LowPt200}.   
For $\pi^{\pm}$ , $K^{\pm}$ and ($p+\bar{p}$) the yields are determined at 
four different ranges of pseudorapidity and  
transverse momentum as was done for the analysis at 62.4~GeV center-of-mass
energy \cite{LowPt624}.  
Both systematic and statistical errors were calculated separately for each measured yield.   
The average combined errors vary from 15\% to 22\% for pions, 17\% to 24\% for kaons and 22\% to 55\% for
($p+\bar{p}$).
The increase of systematic errors for ($p+\bar{p}$) is related to the small number of reconstructed particles 
that leads to significant differences between experimental and Monte Carlo ($p+\bar{p}$) spectra.

In Figure~\ref{comp200_624} the new results are compared with
results from the previous Au+Au  analysis at 62.4 GeV as a function of centrality defined by the number of participating nucleons.
The data for the two first centralities 0-6\% and 6-15\% are merged to match the centrality classes available at the lower energy. The yields shown in Figure~\ref{comp200_624} were also averaged over transverse momentum ranges.
For both energies a similar centrality dependence is observed for $\pi^{\pm}$ , $K^{\pm}$ and ($p+\bar{p}$).
\vspace{-0.2cm}

\vspace{-0.3cm}

It is interesting to compare very low transverse momentum results with the data obtained from the intermediate $p_{T}$ region. 
The comparison is performed using data from the PHENIX experiment \cite{Phenix}  
that has measured positively and negatively charged identified particle spectra in various centrality bins. 
Because the charge sign cannot be 
\vspace{-0.82cm}
\begin{figure}[h!]
\begin{minipage}[hb]{8.3cm}
 \vspace{0.5cm}
 \includegraphics[width=8.3cm]{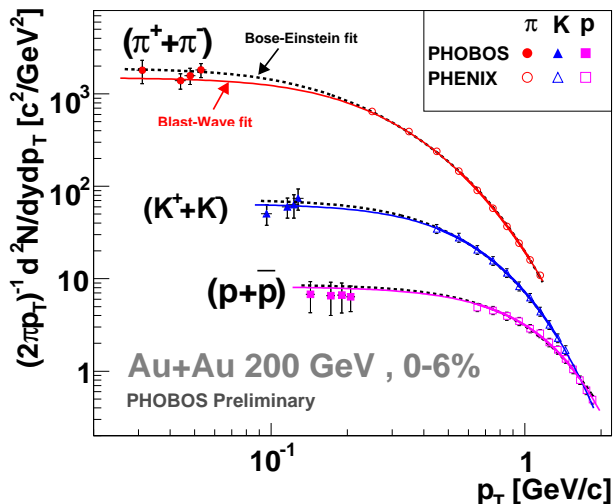}
 \vspace{-1cm}
 \caption{ Blast-Wave (solid~line) and Bose-Einstein (dotted~line) parameterizations fitted over the PHENIX data
           and extrapolated to the lowest $p_{T}$ are compared with PHOBOS results.  }
 \label{BW_BE}    
\end{minipage}
\hfill
\begin{minipage}[hb]{6.9cm}
\vspace{0.45cm}
determined in the analysis of very low $p_{T}$ particles, we compare only the sum of yields of negative and positive charges.
To extrapolate to the lowest $p_{T}$ regions, Blast-Wave \cite{Blast_Wave} and Bose-Einstein \cite{LowPt200} parameterizations are fitted to the PHENIX data (see~Fig.~\ref{BW_BE}).
The Blast-Wave and Bose-Einstein functions fitted in the intermediate $p_{T}$ region are then 
extrapolated to the lowest $p_{T}$.
For the Blast-Wave parameterization $ \beta( r )~=~\beta_{s}( r/R )^{n} $ we assume 
a source radius R = 10 fm and the parameter n is set to 1. 
The Blast-Wave parameters, transverse velocity $\beta_{T}$ and freeze-out temperature $T_{fo}$, are listed in Table~\ref{parameters}.  
\end{minipage}

\end{figure}
\newpage
The Bose-Einstein parameterization, $A\left[ e^{m_{T}/T} \pm 1 \right]^{-1}$  
was fitted to the intermediate $p_{T}$ region and then extrapolated to low $p_{T}$ range.
Temperature parameters $T_{\pi}, T_{k}, T_{p}$ are obtained for each particle type (see Table~\ref{parameters}).
The extrapolation down to low $p_{T}$ is shown in Fig.~\ref{BW_BE} and compared  
with PHOBOS results.
We see that in each centrality bin both extrapolations agree well with our measured data.
No centrality dependence is observed for either Blast-Wave  
parameter, $T_{fo}$ or $\beta_{T}$.
Similar to the previous low $p_{T}$  analysis, no anomalous enhancement of invariant pion yield at very low $p_{T}$ is observed \cite{LowPt200}.
The next important observation is the flattening of the shape of  
particle 
spectra in the very low $p_{T}$ range. 
The flattening clearly visible for ($p+\bar{p}$) increases with particle mass and is consistent 
with a transverse expansion of the strongly-interacting system.

\vspace{-0.3cm}

\begin{table}[htb]
\caption{Parameters obtained from Blast-Wave and Bose-Einstein fits.}
\label{parameters}
\vspace*{0.1cm}
\begin{tabular}{|r|p{2.2cm}|p{2.2cm}||p{2.2cm}|p{2.2cm}|p{2.2cm}|}
\hline
 centrality &  $T_{fo}$ [MeV] & $\beta_{T}$ [c] & $T_{\pi}$ [MeV]   & $T_{k}$ [MeV]  & $T_{p}$ [MeV] \\ \hline
 0-6\%      &  100 $\pm$ 4.0  & 0.81 $\pm$ 0.03 & 229 $\pm$ 5.0 & 291 $\pm$ 11 & 398 $\pm$ 15  \\ \hline 
 6-15\%     &  106 $\pm$ 3.3  & 0.78 $\pm$ 0.03 & 229 $\pm$ 5.0 & 293 $\pm$ 11 & 388 $\pm$ 14  \\ \hline
 15-30\%    &  105 $\pm$ 4.4  & 0.79 $\pm$ 0.03 & 227 $\pm$ 5.0 & 288 $\pm$ 10 & 370 $\pm$ 13  \\ \hline
 30-50\%    &  103 $\pm$ 3.8  & 0.78 $\pm$ 0.03 & 209 $\pm$ 4.8 & 278 $\pm$ 10 & 333 $\pm$ 11  \\ \hline
\end{tabular}
\end{table}

\vspace{0.2cm}

\begin{figure}[h!]

\begin{minipage}[t]{7.2cm}
\includegraphics[width=7cm]{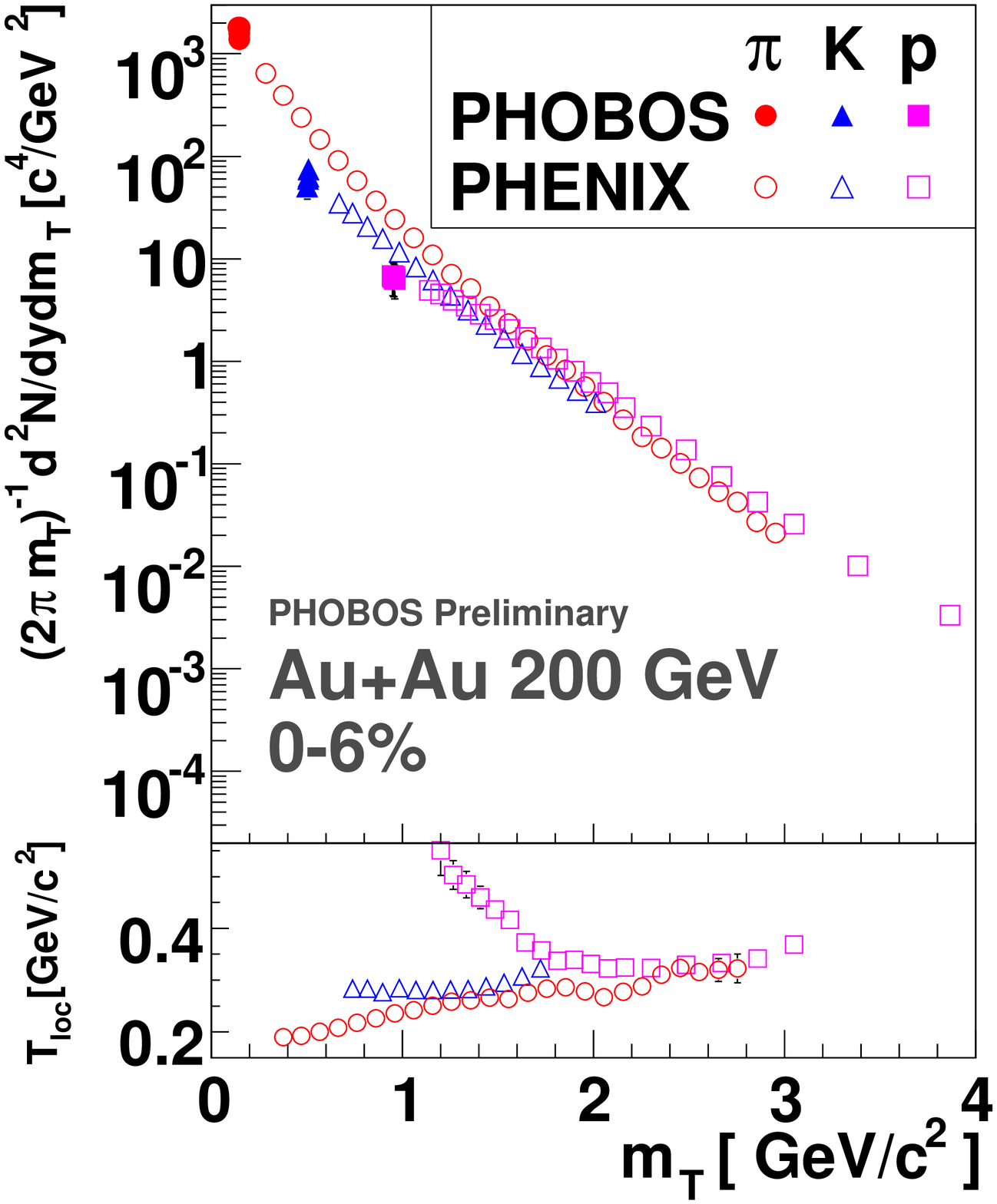} 
\vspace{-0.6cm}
\caption{ $m_T$ spectra and local inverse slope of ($\pi^{+}~+~\pi^{-}$), ($K^{+}~+~K^{-}$) and
         ($p~+~\bar{p}$) at low and intermediate $p_T$ measured in Au+Au at $\sqrt{s_{NN}}$~=~200~GeV. } 
\label{mTscaling_200}
\end{minipage}
\hfill
\begin{minipage}[t]{7.2cm}
\includegraphics[width=7cm]{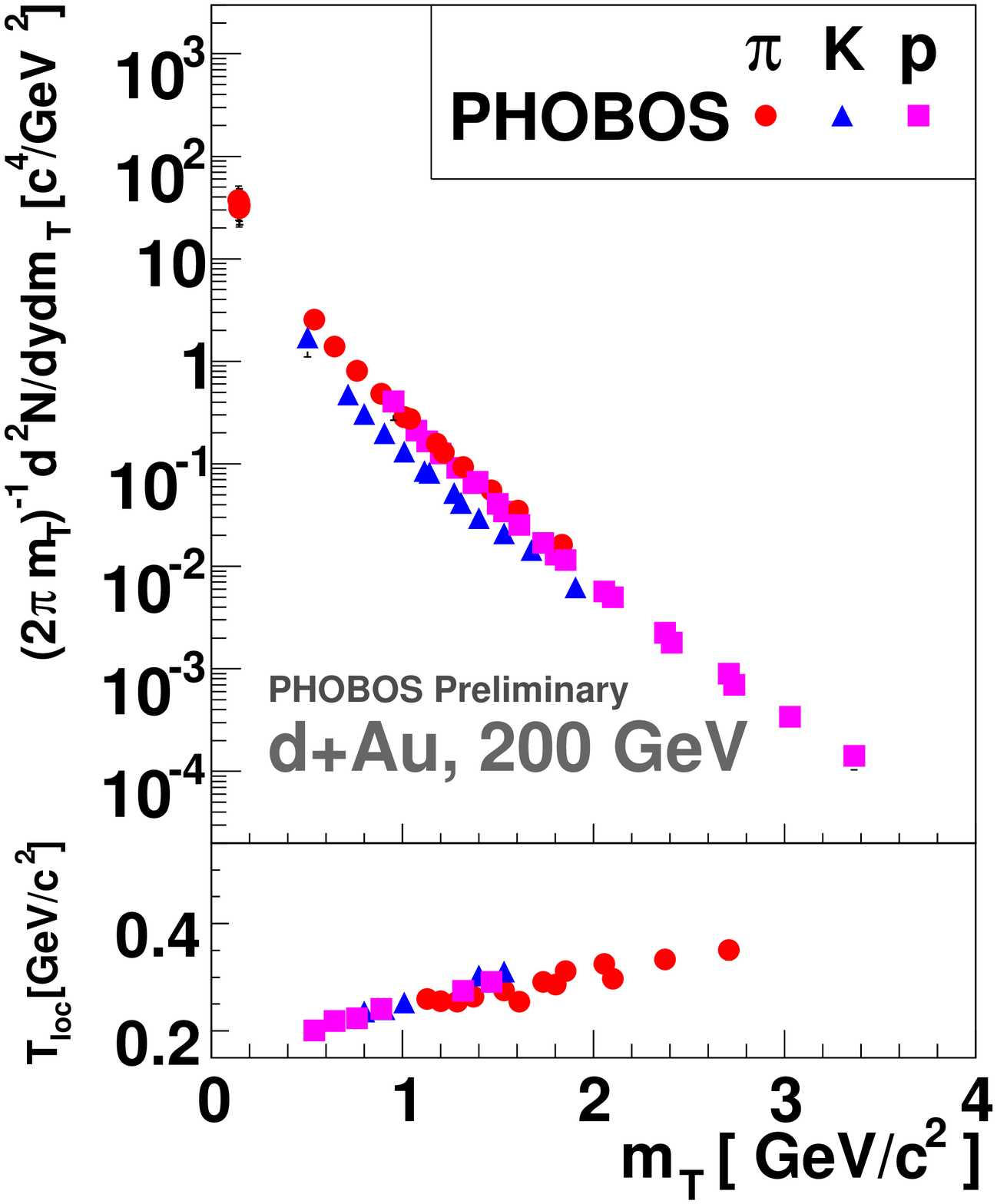} 
\vspace{-0.6cm}
\caption{ $m_T$ spectra and local inverse slope of ($\pi^{+}~+~\pi^{-}$), ($K^{+}~+~K^{-}$) and  
         ($p~+~\bar{p}$) at low and intermediate $p_T$ measured in d+Au at $\sqrt{s_{NN}}$~=~200~GeV. }
\label{mTscaling_624}
\end{minipage}

\end{figure}

We have also investigated the $m_{T}$ spectra ($m_{T} = \sqrt{p_{T}^{2}+m_{0}^{2} }$).   
Figure ~\ref{mTscaling_200} (upper~plot) shows the $m_{T}$ spectra in four centrality classes as well as 
the inverse local slope parameter $T_{loc}$ 
(bottom~plot) calculated in order to more easily show differences 
in the slope of the particle spectra. 
In Au+Au collisions no $m_{T}$ scaling \cite{mTScaling} is observed in the very low $p_{T}$ region,
while d+Au data \cite{dAu_Scaling} at $\sqrt{s_{NN}}$~=~200 GeV (Fig.~\ref{mTscaling_624}) exhibit $m_{T}$ scaling over
the full $p_{T}$ range.

\begin{figure}[h!]
\begin{minipage}[t]{16cm}
\includegraphics[width=16cm]{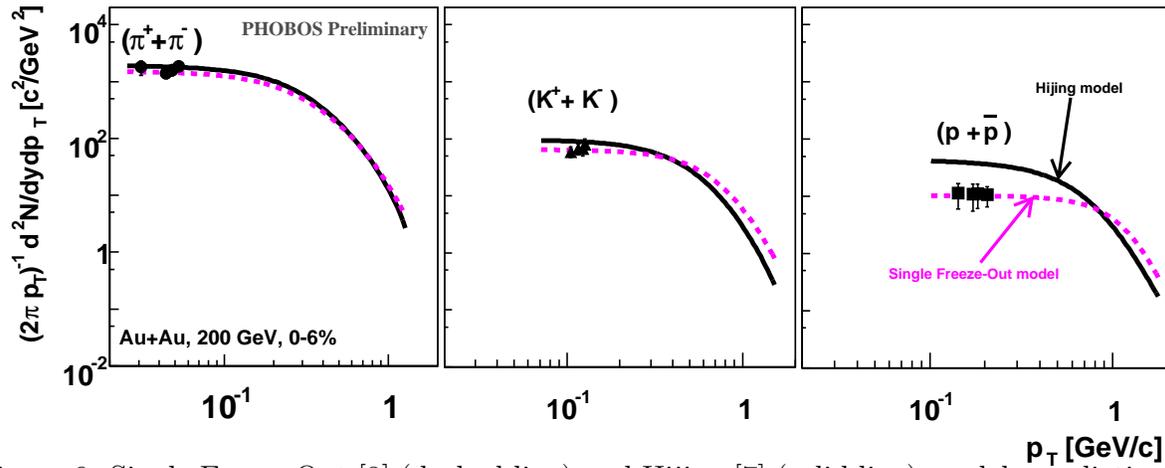} 
\vspace{-1.3cm}
\caption{Single Freeze-Out \cite{FreezeOut} (dashed line) and Hijing \cite{Hijing} (solid line) 
         models predictions compared with very low $p_{T}$ data. }
\label{models}
\end{minipage}

\end{figure}

HIJING \cite{Hijing} and the Single Freeze-Out (Therminator) \cite{FreezeOut} model calculations are compared with low $p_{T}$ data for Au+Au collisions at $\sqrt{s_{NN}}$~=~200 GeV.
The model predictions were obtained for the most central bin separately for $\pi^{\pm}$ , $K^{\pm}$ and ($p+\bar{p}$) (~Fig.~\ref{models}~). 
The comparison shows that both models well describe pion spectra. For heavier particles the single freeze-out model agrees very well with low $p_T$ results.
In contrast, the HIJING model overestimates proton and antiproton yields and the same tendency, although weaker, is seen
in the kaon spectra.

In conclusion, identified particle spectra in Au+Au collisions at $\sqrt{s_{NN}}$ = 200 GeV have been measured at very low $p_{T}$ in 
four centrality bins. Blast-Wave and Bose-Einstein parameterizations well describe the data at very low and intermediate $p_{T}$. 
At the very low transverse momenta spectra, we observe flattening which increases with particle mass consistent with 
rapid transverse expansion of the system.
In Au+Au collisions no $m_T$ scaling is observed in contrast to results from d+Au collisions at $\sqrt{s_{NN}}$ = 200 GeV. The Single Freeze-Out model is consistent with the presented results.

This work was partially supported by U.S. DOE grants 
DE-AC02-98CH10886,
DE-FG02-93ER40802, 
DE-FG02-94ER40818,  
DE-FG02-94ER40865, 
DE-FG02-99ER41099, and
DE-AC02-06CH11357, by U.S. 
NSF grants 9603486, 
0072204,            
and 0245011,        
by Polish MNiSW grant N N202 282234 (2008-2010),
by NSC of Taiwan Contract NSC 89-2112-M-008-024, and
by Hungarian OTKA grant (F 049823).

\vspace{-1.0cm}

\newpage

\end{document}